# Charge migration in metal-organic frameworks

*Maximilian Kriebel,[a,‡] Matthias Hennemann,[a,‡] Frank R. Beierlein,[a] Dana D. Medina-Tautz,[b] Thomas Bein,[b] and Timothy Clark[a]\**

[a] Computer-Chemistry-Center, Department of Chemistry and Pharmacy, Friedrich-Alexander-University Erlangen-Nuremberg, Naegelsbachstr. 25, 91052 Erlangen, Germany.

[b] Department of Chemistry and Center for NanoScience (CeNS), University of Munich (LMU), Butenandtstraße 5-13, 81377 Munich, Germany.

E-mail: Tim.Clark@fau.de

KEYWORDS Metal-organic framework; Charge transport; Semiempirical MO theory; Imaginary time propagation; Conduction paths

ABSTRACT: Charge transport in two zinc metal-organic frameworks (MOFs) has been investigated using periodic semiempirical molecular orbital calculations with the AM1* Hamiltonian. Restricted Hartree-Fock calculations underestimate the band gap (Koopmans theorem), which however becomes more realistic when the wavefunction is allowed to become unrestricted. Charge-transport simulations using propagation of the electron- or hole-density in imaginary time allow charge-transport paths to be determined, although the calculated mobilities must still be improved. In general, the techniques discussed appear to be useful for investigating electroactive MOFs.

## INTRODUCTION

Metal organic frameworks (MOFs) consist of organic ligands and metal clusters connected by chemical bonds. Due to symmetry constraints governing the connectivity of these building blocks, highly porous and crystalline materials are obtained. In combination with high electrical conductivity, MOFs are an interesting case of lightweight yet electrically conducting materials.[1] In this context, MOF-74 is an intriguing framework topology consisting of metal ions (M=Mg, Mn, Fe, Co, Ni, Cu, Zn) connected via phenolate groups of the organic ligand (DOBDC4, 2,5-dihydroxybenzene-1,4-dicarboxylate) forming infinite metal oxide(-M-O-) columns that serve as paths for charge-carrier transport.[2,3,4,5,6,7,8,9] Hopping charge transport has been suggested for the MOF-74 analogs based on band-structure calculations that show narrow valence and conduction bands. Another beneficial property of these frameworks is the exact positioning of the organic ligands as ordered stacks, yielding an open hexagonally shaped pore and hence allowing for the incorporation of electronically complementary molecules into the pore voids for efficient charge separation. In addition, employing photoactive or semiconducting segments as the organic ligands can provide additional possible charge-transfer and -transport routes.[10] Currently, MOF-74 analogs connected via phenolate groups show rather low electrical conductivity in the range of $10^{-11}$-$10^{-13}$ S cm$^{-1}$ with iron as an exceptional case that exhibits electrical conductivity of $10^{-6}$ S cm$^{-1}$.[11] Therefore, studying the possible charge-transport paths and their relation to the organic ligands in these frameworks can help design MOFs that exhibit high electrical conductivity while remaining porous and crystalline. We now present a proof-of-principle study to test the use of large-scale semiempirical molecular orbital (MO) calculations and local properties derived from them to simulate charge transport in zinc MOF74.

## THEORY

MOFs are "soft" crystals, in which disorder may exist; lattice librations are larger than in conventional crystals and enclosed solvent molecules may be very mobile. The MOF structures constructed as described in the Methods section were therefore subjected to equilibration using classical molecular dynamics (MD) and snapshots from the equilibrated phase of the simulations were used for further calculations.

Periodic AM1*[12,13] semiempirical molecular orbital (MO) wavefunctions were used as the basis for calculating local ionization energies[14,15] and local electron affinities.[15,16] These local properties have been used successfully as external potentials for charge-transport using norm-conserving propagation in imaginary time[17] and multi-agent Monte-Carlo[18] techniques, respectively. Propagation of the hole- or excess electron density in imaginary time uses the energy loss in imaginary-time propagation as a surrogate for vibrational and other coupling losses.[17] It has been used in this work in conjunction with a path-search that uses the neighboring cubic layer of lattice points around that in question to find the next low-energy point in the path. The search algorithm did not allow lattice points to be visited twice. The Cartesian directions used throughout are shown in Figure 1.

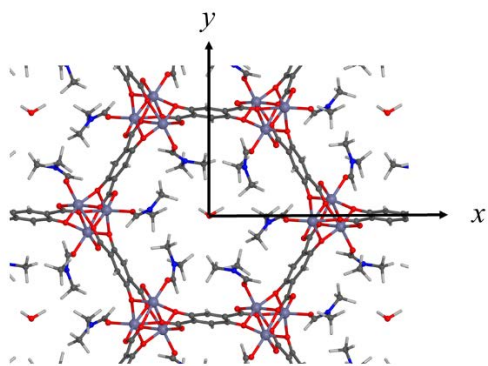

**Figure 1.** the Cartesian directions used throughout. The *z*-direction is perpendicular to the plane of the picture.

**RESULTS**

**Molecular-dynamics simulations**

The classical geometry optimizations and MD simulations show that the simulated systems are stable. Cell dimensions and atom positions of the organic linkers, the -O-Zn-O- network and dimethyl-formamide (DMF) ligands hardly change after geometry optimization and equilibration. The water molecules rotate continuously but only rarely swap positions. "Soft" parameters such as the inter-stack distances oscillate within a restricted range.

**Local-property maps**

Figure 2 shows the isovalue surfaces for the local ionization energy, $IE_L$ (contour level = 24 eV) and the local electron affinity (contour level = -1 eV) for a zinc-MOF74 snapshot from the simulation.

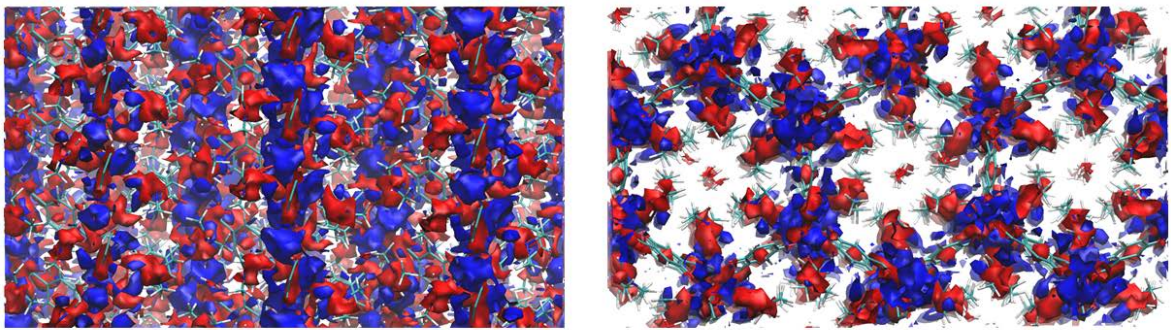

**Figure 2.** Side and top views of the isovalue surfaces of $IE_L$ (red, contour level = 24 eV) and $EA_L$ (blue, contour level = -1 eV) for zinc-MOF74 calculated with AM1*.

Both local properties show relatively continuous paths along the zinc stacks. The columns are largely isolated from each other by the bridging ligands and no obvious conduction paths are evident perpendicular to the stack direction. This contrasts with the picture found for zinc-ANMOF74, as shown in Figure 3.

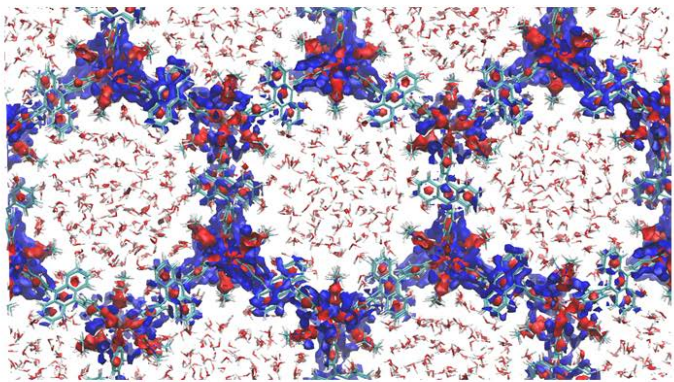

**Figure 3:** Side and top views of the isovalue surfaces of $IE_L$ (red, contour level = 24 eV) and $EA_L$ (blue, contour level = -1 eV) for zinc-ANMOF74 calculated with AM1*.

The anthracene moieties in the bridging ligands are more electroactive (have a lower ionization potential and higher electron affinity than the phenyl linkers) and therefore begin to play a role in developing conjugation between the metal stacks. In particular, the side view shows that, in certain geometrical arrangements that are accessible in the simulations, distinct inter-stack conjugation over stacks of two anthracene units can occur. This picture suggests that increasing the electron affinity of the linker even more might result in an alternative electron-transport path geometrically distinct from that for hole-conduction.

**Density of states**

Figure 4 shows the Koopmans theorem density of states for zinc-MOF74 and Figure 5 that for zinc-ANMOF74. The vertical scales are the same in the two figures, which were constructed by adding Gaussians of half-width 0.1 eV centered at each orbital eigenvalue. In both cases, the calculations indicate a band gap of approximately 2 eV.

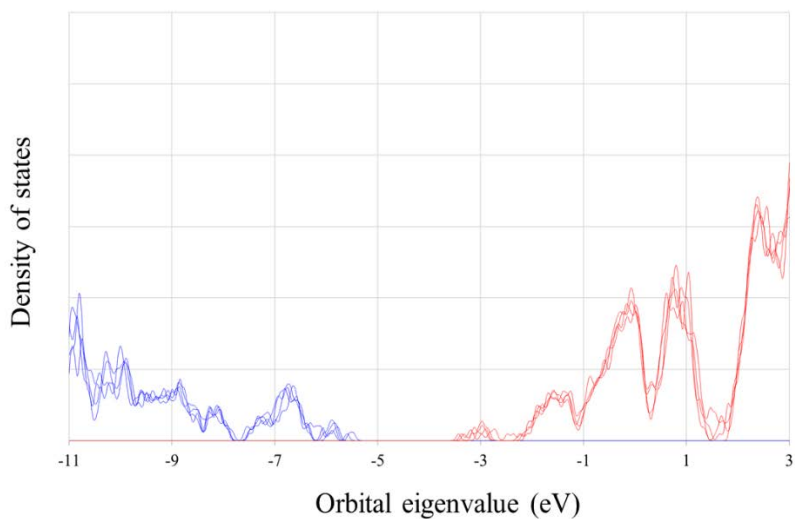

**Figure 4:** Koopmans theorem density of states diagram for zinc-MOF74 calculated with AM1*. The diagram was constructed as described in the text.

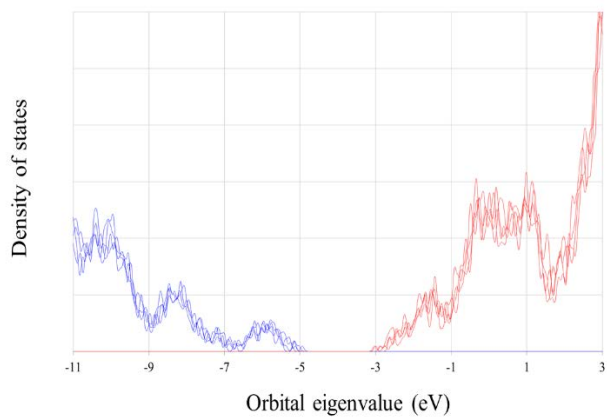

**Figure 5:** Koopmans theorem density of states diagram for zinc-ANMOF74 calculated with AM1*. The diagram was constructed as described in the text.

Because, however, structures with so many aromatic moieties exhibit RHF → UHF instability within NDDO-based theory, we also calculated the UHF natural orbitals[19] for a zinc-MOF74 snapshot. The results are shown in Figure 6.

The UHF natural orbital (UNO) occupation numbers show clear open-shell character, as expected.[19] Above and below the band gap, which is now slightly higher than 4 eV, the UNOs are essentially singly occupied. This indicates that UNO-CI calculations,[19] which are not yet implemented for periodic systems in EMPIRE, should give reliable band gaps.

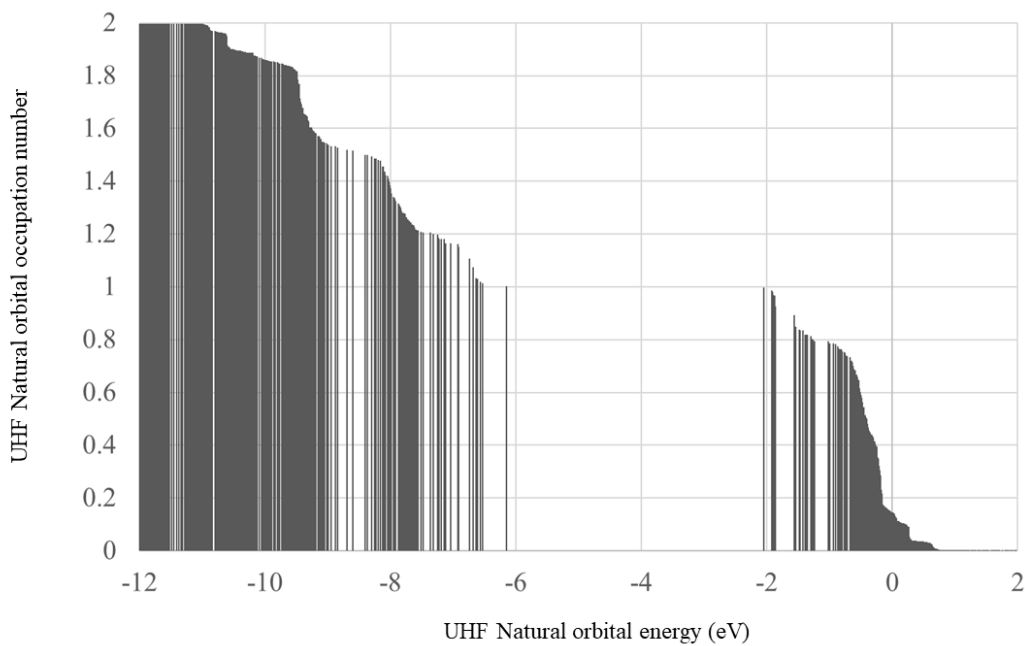

**Figure 6:** UHF Natural orbital occupation numbers for zinc-AMOF74 calculated with AM1*.

**Charge propagation simulations: Mobilities**

Mobility values given by the imaginary time technique are neither comparable between different systems (MOF74 and ANMOF74), nor between hole- and electron conductance. We therefore $z$-direction for each system, as shown in Table 1.

**Table 1.** Calculated relative mobilities for electrons and holes in the three Cartesian directions for zinc-MOF74 and zinc-ANMOF74.

| direction | Zinc-MOF74 | | Zinc-ANMOF74 | |
| --- | --- | --- | --- | --- |
| | electron | hole | electron | hole |
| $x$ | 96.6% | 99.7% | 82.3% | 87.4% |
| $y$ | 89.1% | 93.7% | 78.3% | 89.2% |
| $z$ | 100.0% | 100.0% | 100.0% | 100.0% |

The differentiation between the $z$- and $x$-directions is almost non-existent for zinc-MOF74, which suggests that the calculated mobilities are not reliable. Zinc-ANMOF74, however, shows a clearer preference for the $z$-direction. Clearly, calculation of mobilities within the virtual-time propagation technique must be improved.

**Charge propagation simulations: Conductance paths**

Conductance paths were extracted from the simulation trajectories using a simple search in which lattice points can only be visited once and each move takes place to the energetically lowest neighbor. The results for the simulations in the three Cartesian directions are shown in Figure 7.

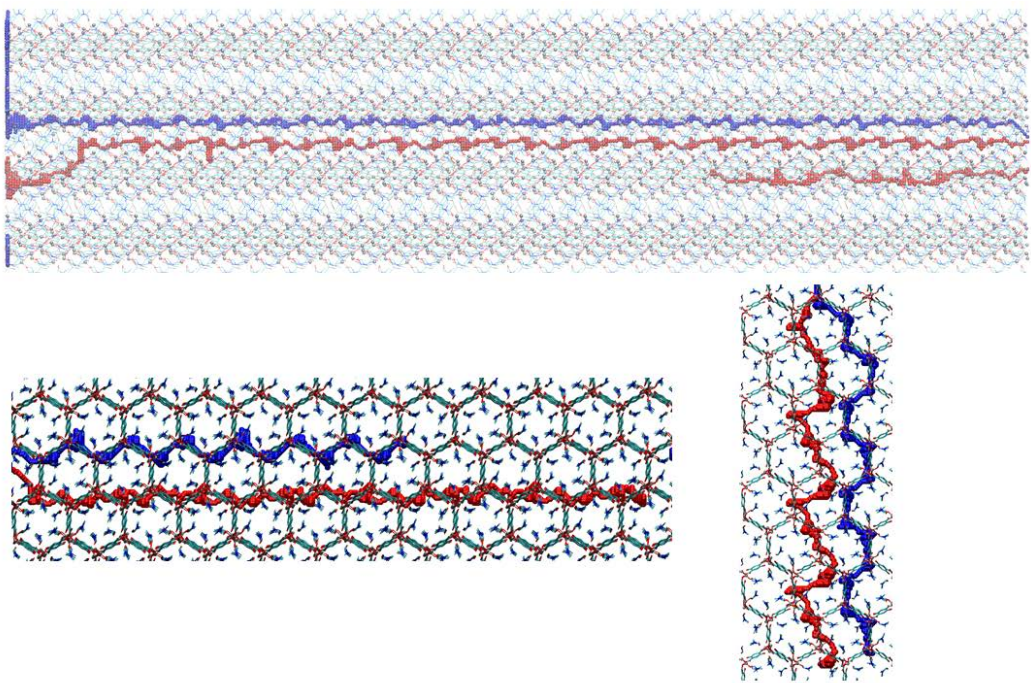

**Figure 7:** Hole- (red) and electron- (blue) conduction paths in zinc-MOF74. The DMF and water molecules have been omitted for clarity. Top, *z*-direction; bottom left, *y*-direction; bottom right, *x*-direction.

The expected paths in the *z*-direction (top) follow the ZnO-stacks closely, although the hole-path occasionally deviates in the perpendicular direction. The (less favorable) paths in the *y*- and *x*-directions (bottom left and right, respectively) differ fairly strongly from each other. The electron-path follows the framework of the linkers closely, whereas the hole-path makes use of especially the coordinated DMF molecules.

The situation is different for zinc-ANMOF74, as shown in Figure 8. The linking anthracene linkers play a far more active role in the conductance paths, particularly for hole-conductance, than the phenyl linkers in zinc-MOF74. Once again, hole-conductance follows the molecular network closely, whereas hole-conductance also involves the coordinated DMF molecules and often runs parallel to the network linkages. This can be seen particularly clearly for hole-conductance in the

*x*-direction, which proceeds almost in straight lines between the ZnO columns, whereas electron-conduction paths zigzag with the MOF framework.

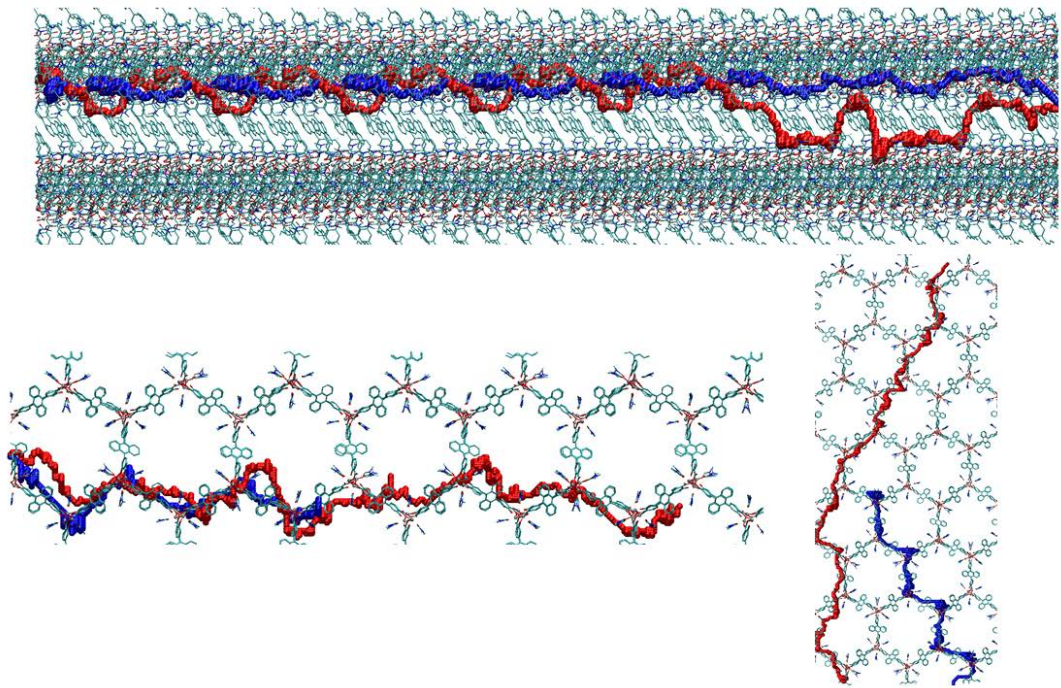

**Figure 8:** Hole- (red) and electron- (blue) conduction paths in zinc-ANMOF74. The DMF and water molecules have been omitted for clarity. Top, *z*-direction; bottom left, *y*-direction; bottom right, *x*-direction.

## METHODS

The structure of MOF-74 was first constructed from .cif files based on experimental results.[2] Partially occupied water positions and disordered DMF were replaced by the correct proportions of the alternative structures. AnMOF was constructed in analogy to the experimental MOF-74 structure, including DMF ligands and water. Hexagonal simulation cells of dimensions $25.9 \times 25.9 \times 27.3$ Å (MOF-74) and $46.3 \times 46.3 \times 23.1$ Å (AnMOF), respectively, were constructed by stacking four unit cells of each. The UFF force field[20] with AM1[21,22] Coulson charges was used for the classical molecular dynamics (MD) simulations. System preparation was performed using slightly modified tools developed by Boyd et al.[23,24] A cutoff of 10 Å was used for the nonbonding interactions; Coulomb interactions beyond this distance were treated using a particle-particle Ewald approach.[25,26] After initial geometry optimization, the system was equilibrated for 11 ns (1 ns heat-up using a Langevin thermostat in the NVE ensemble and 10 ns NVT (Nosé-Hoover thermostat) simulation, 1 fs timestep, 298 K) using LAMMPS version 12 Dec 2018.[27,28] Snapshots taken from the following 40 ns of the NVT simulations were used for further processing. Visualization was performed using Vipster v1.17b[29] and VMD 1.9.3.[30,31]

Semiempirical MO-calculations using the AM1*[12,13] Hamiltonian within the restricted Hartree-Fock formalism were carried out using the cluster version of EMPIRE.[32,33,34] No local approximations were used. AM1* differs from other common semiempirical Hamiltonians in that it includes an occupied *3d*-shell of electrons on Zn, rather than including it in the frozen core.[13] The periodic[33] AM1* calculations were performed on the simulation cells consisting of 1620 atoms (MOF-74) and 4740 atoms (AnMOF) within hexagonal cells of 24.9 × 24.9 × 25.0 Å (MOF-74) and 46.2 × 46.2 × 25.1 Å (AnMOF), respectively.

Three-dimensional local-ionization energy[14] and local electron affinity[15,16] maps were generated with the EH5cube program[35] and stored in Gaussian cube-file format[36] as orthogonal cells of 24.9 × 43.2 × 25.0 Å (MOF-74) and 46.2 × 80.0 × 25.1 Å (AnMOF), respectively.

Propagation of the hole- or electron density in imaginary time used an in-house program,[17] as did the conversion of the results into conductance paths. Simulations of charge-transport in the different Cartesian directions were performed by applying a bias in the appropriate direction to the external potential, as described in reference 17.

AUTHOR INFORMATION

**Corresponding Author**

* Tim.Clark@fau.de

**Author Contributions**

The manuscript was written through contributions of all authors. All authors have given approval to the final version of the manuscript. ‡These authors contributed equally.

**Funding Sources**

This work was supported by the Deutsche Forschungsgemeinschaft within SPP1928 "Coordination Networks: Building Blocks for Functional Systems."

ACKNOWLEDGMENT

The authors wish to thank Sebastian Gsänger for help with the Lammps MD simulations and the Vipster software.

REFERENCES

[1] Medina, D. D.; Mähringer, A.; Bein, T. Electroactive metalorganic frameworks. *Isr. J. Chem.* **2018**, *58*, 1089-1101.

[2] Rosi, N. L.; Kim, J.; Eddaoudi, M.; Chen, B.; O'Keeffe, M.; Yaghi, O. M. Rod packings and metal−organic frameworks constructed from rod-shaped secondary building units. *J. Am. Chem. Soc.* **2005**, *127*, 1504-1518.

[3] Dietzel, P. D. C.; Morita, Y.; Blom, R.; Fjellvåg, H. An in situ high-temperature single-crystal investigation of a dehydrated metal-organic framework compound and field-induced magnetization of one-dimensional metal-oxygen chains. *Angew. Chem. Int. Ed.* **2005**, *44*, 6354-6358.

[4] Dietzel, P. D. C.; Panella, B.; Hirscher, M.; Blom, R.; Fjellvåg, H. Hydrogen adsorption in a nickel based coordination polymer with open metal sites in the cylindrical cavities of the desolvated framework. *Chem. Commun.* **2006**, 959-961.

[5] Dietzel, P. D. C.; Blom, R.; Fjellvåg, H. Base-induced formation of two magnesium metal-organic framework compounds with a bifunctional tetratopic ligand. *Eur. J. Inorg. Chem.* **2008**, 3624–3632.

[6] Zhou, W.; Wu, H.; Yildirim, T. Enhanced $H_2$ adsorption in isostructural metal−organic frameworks with open metal sites: Strong dependence of the binding strength on metal ions. *J. Am. Chem. Soc.* **2008**, *130*, 15268-15269.

[7] Bloch, E. D.; Murray, L. J.; Queen, W. L.; Chavan, S.; Maximoff, S. N.; Bigi, J. P.; Krishna, R.; Peterson, V. K.; Grandjean, F.; Long G. J.; Smit, B.; Bordiga, S.; Brown, C. M.; Long, J. R. Selective binding of $O_2$ over $N_2$ in a redox–active metal–organic framework with open iron(II) coordination sites. *J. Am. Chem. Soc*. **2011**, *133*, 14814-14822.

[8] Sanz, R.; Martinez, F.; Orcajo, G.; Wojtas, L.; Briones, D. Synthesis of a honeycomb-like Cu-based metal–organic framework and its carbon dioxide adsorption behavior. *Dalton Trans.* **2013**, *42*, 2392-2398.

[9] Sun, L. Hendon, C. H.; Minier, M. A.; Walsh, A.; Dincă, M. Million-Fold Electrical Conductivity Enhancement in Fe2(DEBDC) versus Mn2(DEBDC) (E = S, O). *J. Am. Chem. Soc.* **2015**, *137*, 6164–6167.

[10] Foster, M. E. Azoulay, J. D. Wong, B. M.; Allendorf, M. D. Novel metal-organic framework linkers for light harvesting applications. *Chem. Sci.* **2014**, *5*, 2081-2090.

[11] Sun, L.; Campbell, M. G.; Dincă, M. Electrically conductive porous metal–organic frameworks. *Angew. Chem. Int. Ed.* **2016**, *55*, 3566-3579

[12] Winget, P.; Horn, A. H. C.; Selçuki, C.; Martin, B.; Clark, T. AM1* parameters for phosphorous, sulfur and chlorine. *J. Mol. Model.* **2003**, *9*, 408-414.

[13] Kayi, H.; Clark, T. AM1* Parameters for copper and zinc. *J. Mol. Model.*, **2007**, *13*, 965-979.

[14] Sjoberg P.; Murray J. S.; Brinck T.; Politzer P. A. Average local ionization energies on the molecular surfaces of aromatic systems as guides to chemical reactivity. *Can. J. Chem.* **1990**, *68*, 1440-1443.

[15] Ehresmann, B.; de Groot, M. J.; Alex, A.; Clark, T. New molecular descriptors based on local properties at the molecular surface and a boiling-point model derived from them. *J. Chem. Inf. Comp. Sci.* **2004**, *44*, 658-668.

[16] Clark, T.; The local electron affinity for non-minimal basis sets, *J. Mol. Model.* **2010**, *16*, 1231-1238.

[17] Kriebel, M.; Sharapa, D.; Clark, T. Charge transport in organic materials: Norm-conserving imaginary time propagation with the local ionization energy as external potential. *J. Chem. Theor. Comput.* 2017, 13, 6308–6316.

[18] Bauer, T.; Jäger, C. M.; Jordan, M. J. T.; Clark, T. A multi-agent quantum Monte-Carlo model for charge transport: Application to organic field-effect transistors. J. Chem. Phys. 2015, 143, 044114.

[19] Dral, P. O.: Clark, T. Semiempirical UNO-CAS and UNO-CI: Method and applications in nanoelectronics, *J. Phys. Chem. A*, **2011**, *115*, 11303-11312.

[20] Rappe, A. K.; Casewit, C. J.; Colwell, K. S.; Goddard, W. A.; Skiff, W. M. Uff, a Full Periodic Table Force Field for Molecular Mechanics and Molecular Dynamics Simulations. *J. Am. Chem. Soc.* **1992,** *114* (25), 10024-10035.

[21] Dewar, M. J. S.; Zoebisch, E. G.; Healy, E. F.; Stewart, J. J. P. AM1 - A new general purpose quantum mechanical molecular model. *J. Am. Chem. Soc.* **1985**, *107*, 3902-3909.

[22] Holder, A. J. AM1 in *Encyclopedia of Computational Chemistry*, Schleyer, P. v. R.; Allinger, N. L.; Clark, ;T. Gasteiger, J.; Kollman, P. A.; Schaefer III, H. F.; Schreiner, P. R. (Eds.), Wiley, Chichester, 1998, Vol. 1, pp. 8-11.

[23] Boyd, P. G.; Moosavi, S. M.; Witman, M.; Smit, B. Force-Field Prediction of Materials Properties in Metal-Organic Frameworks. *J. Phys. Chem. Lett.* **2017,** *8* (2), 357-363.

[24] Lammps_Interface on Github. https://github.com/peteboyd/lammps_interface (accessed 20. May 2019).

[25] Hockney, R. W.; Eastwood, J. W. *Computer Simulation Using Particles*. Adam Hilger: NY 1989.

[26] Darden, T.; York, D.; Pedersen, L. Particle Mesh Ewald: An N Log(N) Method for Ewald Sums in Large Systems. *J. Chem. Phys.* **1993,** *98*, 10089-10092.

[27] Plimpton, S. Fast Parallel Algorithms for Short-Range Molecular Dynamics, *J. Comp. Phys.* **1995**, *117*, 1-19.

[28] http://lammps.sandia.gov accessed 17th May 2019.

[29] Vipster on Github. https://sgsaenger.github.io/vipster/ (accessed 17. May 2019).

[30] http://www.ks.uiuc.edu/Research/vmd/ (accessed 17. May 2019).

[31] Humphrey, W.; Dalke, A.; Schulten, K. Vmd - Visual Molecular Dynamics. *J. Molec. Graphics* **1996,** *14*, 33-38.

[32] Hennemann, M.; Clark, T. EMPIRE: A highly parallel semiempirical molecular orbital program: 1: Self-consistent field calculations. *J. Mol. Model.* **2014**, *20*, 2331.

[33] Margraf, J. T.; Hennemann, M.; Meyer, B.; Clark, T. EMPIRE: A highly parallel semiempirical molecular orbital program: 2: Periodic boundary conditions. *J. Mol. Model.* **2015**, *21*, 144.

[34] Hennemann, M.; Margraf, J. T.; Meyer, B.; Clark, T. EMPIRE19, Cepos InSilico GmbH, Obermichelbach, **2019**. http://www.ceposinsilico.de/products/empire.htm

[35] Hennemann, M. EH5cube, Cepos InSilico GmbH, Obermichelbach, **2019**.

[36] https://gaussian.com/cubegen/ accessed 16. May 2019.